# Optical generation of a spatially variant two-dimensional lattice structure by using a phase only spatial light modulator

Manish Kumar[a)] and Joby Joseph[b)]

*Photonics Research Laboratory, Department of Physics, Indian Institute of Technology Delhi, New Delhi-110016, India*

We propose a simple and straightforward method to generate a spatially variant lattice structures by optical interference lithography method. Using this method, it is possible to independently vary the orientation and period of the two-dimensional lattice. The method consists of two steps which are: numerical synthesis of corresponding phase mask by employing a two-dimensional integrated gradient calculations and experimental implementation of synthesized phase mask by making use of a phase only spatial light modulator in an optical 4f Fourier filtering setup. As a working example, we provide the experimental fabrication of a spatially variant square lattice structure which has the possibility to guide a Gaussian beam through a 90° bend by photonic crystal self-collimation phenomena. The method is digitally reconfigurable, is completely scalable and could be extended to other kind of lattices as well.

Photonic crystals are the ultimate structured materials with the ability to control the flow of light[1] which leads to so many applications. But perfectly periodic photonic crystals have limited functionality and it becomes essential to have more degrees of freedom in photonic crystal design to make way for more devices. More degrees of freedom could be added by spatially varying some of the properties of photonic crystal i.e. lattice orientation, lattice spacing and fill-factor[2-4] etc. A single step, low cost, large area fabrication of periodic (or quasi periodic) photonic crystal structures is possible by the method of interference lithography[5,6] and few recent advances with phase only spatial light modulator (SLM) assisted interference lithography has enabled the fabrication of some unconventional photonic crystal structures. These structures include embedding of defect sites[7-9], creation of line defects[10], dual lattice structures[11] and gradient structures[12,13].

---

[a)] Author to whom correspondence should be addressed. Electronic mail: manishk.iitd@gmail.com
[b)] Electronic mail: joby@physics.iitd.ac.in

This approach is better than other fabrication approaches like e-beam lithography[14] or direct laser writing[15] because they being serial in nature are prone to errors and are time consuming. So, it would be very desirable if this SLM assisted interference lithography approach could be extended to include the spatially variant lattice structures as well.

In this work, we present a method for generating spatially variant optical lattice wave-field towards the fabrication of spatially variant lattice structures. We report a single step optical generation and fabrication of spatially variant optical lattice wave-field and lattice structures. This method enables simultaneous and independent control over the lattice orientation and lattice period. One of such spatially variant lattices has recently been studied[4,16]. In the study in ref.4, a method for direct computation/modeling of physical geometry of a spatially variant lattice has been proposed which was then fabricated by 3D printing based serial writing process in ref. 16. In our method, we extend the numerical synthesis process of spatially variant lattice to extract a phase only mask which is then displayed, using a phase only SLM, in an optical 4f Fourier filtering setup for direct single step optical generation and fabrication of corresponding 2D spatially varying lattice wave-field and structure. Although the phase extraction process is inspired by the method of phase only encoding of periodic lattice wave-fields[17,18], the actual phase synthesis process for spatially varying lattice is not that straight forward. We modify the phase synthesis process based on an integrated gradient calculation of spatially varying wave-vector lattice **K**. The phase only mask generated by our method is used for direct fabrication of calculated lattice in a photosensitive medium. Thus, our approach consists of two parts: numerical synthesis of the spatially variant phase mask and experimental implementation of the synthesized phase mask by displaying it on a phase only SLM which is employed in a simple 4f Fourier filtering setup.

In this letter, we start with the details of the synthesis of phase mask for a spatially variant lattice wave-field and then describe its experimental implementation. We describe how a phase mask could be synthesized for varying lattice orientation and periodicity parameters for the case of a 2D square lattice. As an example, we take a spatially varying square lattice which changes in the lattice orientation parameter as per the azimuthal angle. This lattice is important as it has been shown to guide light through a 90° bend[4,16] by self-collimation phenomena[19] and could be a very useful photonic element. To understand the phase mask synthesis process, we start with the case of a multiple beam interference wave-field where all the beams are spread symmetrically along the surface of a cone making same tilt angle with the normal. Such a multiple (say $n$ in number) beam interference wave-field is mathematically represented by:

$$\mathbf{E}_{Res}(\mathbf{r}) = \sum_{j=1}^{n} \mathbf{E}_j e^{i(\mathbf{k}_j \cdot \mathbf{r})}. \tag{1}$$

Here the wave-vector $\mathbf{k}_j$ could be expressed in terms of the tilt angle, $\theta$, it makes with normal to the interference plane by:

$$\mathbf{k}_j = k \times [\{\cos(q_j\pi) \times \sin\theta\}\hat{x} + \{\sin(q_j\pi) \times \sin\theta\}\hat{y} + \cos\theta\hat{z}], \tag{2}$$

where $k = 2\pi/\lambda$ ($\lambda$ being the wavelength of laser in air) and coefficient $q_j = 2(j-1)/n$.

The corresponding kinoform (phase only component) of such a wave-field (Eq. 1) is found to be self-sufficient to help encode the original wave-field efficiently through an optical 4f filtering process[8-11]. For a uniform square lattice wave-field, $n = 4$ in above equation and we notice that it is effectively formed by superposition of four plane waves which have the same tilt $\theta$ and are symmetrically distributed along +x, -x, +y and –y axis. The resultant lattice wave-field, formed due to interference of such four unit amplitude plane waves, could be represented by:

$$\mathbf{E}_{Res}(\mathbf{r}) = \exp(i\varphi_x) + \exp(i\varphi_y) + \exp(-i\varphi_x) + \exp(-i\varphi_y), \tag{3}$$

where $\varphi_x$ and $\varphi_y$ are the phase functions and are given by a dot product between the beam wave-vector ($\mathbf{k}_j$) and position vector ($\mathbf{r}$). So, $\varphi_x = \mathbf{k}_1.\mathbf{r} = k(\sin\theta)x = -k(\cos(\pi)\sin\theta)x = -\mathbf{k}_3.\mathbf{r}$ and $\varphi_y = \mathbf{k}_2.\mathbf{r} = k(\sin(\pi/2)\sin\theta)y = -k(\sin(3\pi/2)\sin\theta)y = -\mathbf{k}_4.\mathbf{r}$. It is easy to notice that $\varphi_x$ and $\varphi_y$ are related through a simple matrix transpose (or a 90° anticlockwise rotation in a geometrical sense).

In order to extend this process from uniform to spatially variant lattice, we need to include the lattice orientation and lattice spacing/period parameters which control or define the spatially variant lattice. Fill-factor as a third parameter may give an additional control over the resultant crystal structure. At first, it may appear that it is easy to incorporate the lattice orientation and lattice spacing parameters by synthesizing a spatially variant wave-vector lattice $\mathbf{K}$ and then directly calculate the corresponding phase contributions ($\varphi$) by making use of dot product of synthesized $\mathbf{K}$ lattice with position vector as noted in previous paragraph. Fig. 1 shows the result of such an exercise for a simple case of 1D spatially varying grating structure where corresponding 1D interference profile at each stage is calculated by taking absolute square of the resultant total field $\mathbf{E}_{Res}$ = exp($i\varphi$) + exp($-i\varphi$). The mathematical construction of spatially varying $\mathbf{K}$ lattice by including lattice orientation parameter ($\Theta$) and lattice spacing parameter ($P$) could be summarized by the following simple equation,

$$\mathbf{K}_j(\mathbf{r}) = [\cos(\phi_j + \Theta(\mathbf{r}))\hat{\mathbf{x}} + \sin(\phi_j + \Theta(\mathbf{r}))\hat{\mathbf{y}}] \times 2\pi / (\lambda \times P(\mathbf{r})),  \quad (4)$$

where $\phi_j = \cos^{-1}(\mathbf{k}_j \cdot \hat{x})$. $\quad (5)$

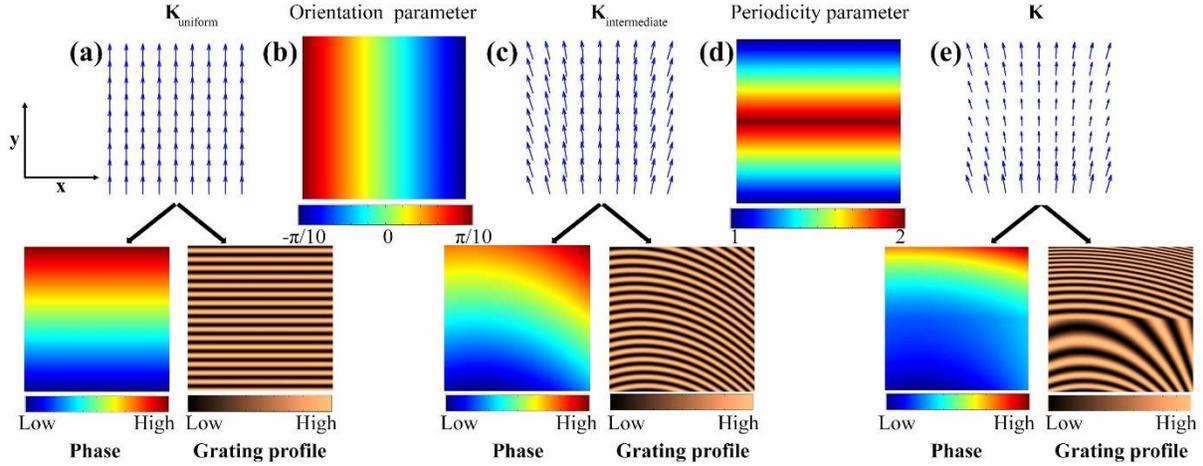

FIG. 1: (Essential Color) Direct construction of 1D spatially varying grating by making use of spatially varying $\mathbf{K}$ lattice. (a) uniform $\mathbf{K}$ lattice, (b) spatially varying orientation parameter (in radian), (c) intermediate $\mathbf{K}$ lattice resulting from application of the orientation parameter on uniform $\mathbf{K}$ lattice, (d) spatially varying periodicity parameter (relative scaling factor), (e) final spatially varying $\mathbf{K}$ lattice resulting after applying periodicity parameter to intermediate $\mathbf{K}$ lattice. The figures in lower row show direct construction of phase variation and grating structure corresponding to each of the $\mathbf{K}$ lattices.

For the case of square lattice, $\phi_j$ gets simplified to $(j-1)\times\pi/2$. We notice from the results in Fig. 1 that a direct construction of the grating from the spatially varying $\mathbf{K}$ function does not render it correctly/smoothly. It is so because the spatially varying $\mathbf{K}$ function is related to the gradient of phase function implying, the correct method to calculate phase functions $\varphi_x$ and $\varphi_y$ is by making use of integrated gradient[4], as explained below. In fact, this problem is similar to the problem of wave-front estimation from a given set of measured wave-front slope data which has been widely encountered in the field of adaptive optics[20]. We do the integrated gradient calculation by a numerical method where the derivatives are approximated by finite difference method and then the generated large set of equations are solved numerically in the least square sense[21]. For the case of square lattice we need to calculate only two phase functions i.e. $\varphi_x$ and $\varphi_y$. The corresponding gradient equations to be solved are

$$\nabla \varphi_x(\boldsymbol{r}) = \mathbf{K}_1(\boldsymbol{r}) = [\cos(\Theta(\boldsymbol{r}))\hat{\mathbf{x}} + \sin(\Theta(\boldsymbol{r}))\hat{\mathbf{y}}] \times 2\pi / (\lambda \times P(\boldsymbol{r})) \qquad (6)$$

and,

$$\nabla \varphi_y(\boldsymbol{r}) = \mathbf{K}_2(\boldsymbol{r}) = [-\sin(\Theta(\boldsymbol{r}))\hat{\mathbf{x}} + \cos(\Theta(\boldsymbol{r}))\hat{\mathbf{y}}] \times 2\pi / (\lambda \times P(\boldsymbol{r})) . \qquad (7)$$

Once $\varphi_x$ and $\varphi_y$ have been calculated by numerically solving the above equations, it is easy to obtain the resultant 2D square lattice wave-field by

$$\mathbf{E}_{\text{Res}}(\boldsymbol{r}) = e^{i\varphi_x(\boldsymbol{r})} + e^{i\varphi_y(\boldsymbol{r})} + e^{-i\varphi_x(\boldsymbol{r})} + e^{-i\varphi_y(\boldsymbol{r})} = \left[e^{i\varphi_x(\boldsymbol{r})} + e^{-i\varphi_x(\boldsymbol{r})}\right] + \left[e^{i\varphi_y(\boldsymbol{r})} + e^{-i\varphi_y(\boldsymbol{r})}\right]. \qquad (8)$$

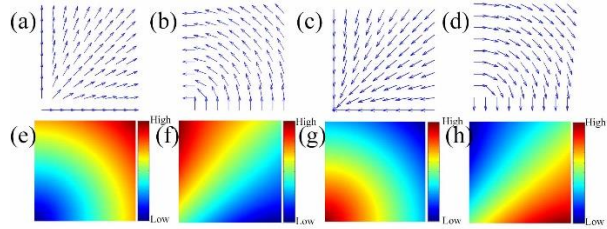

FIG. 2: (Essential Color) Synthesis of spatially varying phase functions by making use of integrated gradient calculations. (a)-(d) are spatially varying $\mathbf{K}_1$, $\mathbf{K}_2$, $\mathbf{K}_3$ and $\mathbf{K}_4$ lattices due to application of azimuthally varying orientation function. (e)-(h) are corresponding numerically calculated phase functions $\varphi_1$, $\varphi_2$, $\varphi_3$, and $\varphi_4$ respectively.

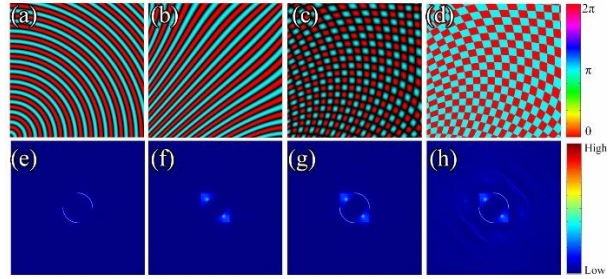

FIG. 3: (Essential Color) Construction of spatially variant lattice wave-field and corresponding phase profile. (a) 1D spatially variant lattice wave-field due to $\varphi_1$ and $\varphi_3$ phase contribution, (b) 1D spatially variant lattice wave-field due to $\varphi_2$ and $\varphi_4$ phase contributions, (c) 2D spatially variant lattice wave-field by adding both 1D gratings, (d) phase only component of 2D spatially variant lattice wave-field, (e)-(h) Numerically obtained Fourier transform of the complex functions represented by (a)-(d) respectively.

The two terms on the right hand side of Eq. 8 construct two 1D gratings which upon a coherent superposition give the desired spatially varying square lattice. Fig. 2 displays the construction of the four phase functions for the case of a spatially varying lattice where the lattice orientation changes from 0 to π as per the azimuth angle. In Fig. 3 we show the construction of corresponding spatially varying 1D gratings and square lattice wave-field by making use of Eq. 8. The numerically calculated Fourier transforms of these wave-fields are also shown which gives a better insight in the nature of wave-front required for formation of this spatially varying lattice. Fig. 3 shows that the present case of spatially varying lattice wave-field is much similar to the case of discrete non-diffracting wave-field as the phase only component of this wave-field also gives rise to some higher order diffraction terms which are non-overlapping with desired first order diffraction term. This makes it experimentally feasible to do a complete encoding of spatially varying lattice wave-field by its kinoform[18]. Although we have considered a simple spatially varying lattice involving the variation of lattice orientation alone, a more complex lattice which involves a variation in periodicity as well could be encoded in exactly similar fashion provided the diffraction orders of corresponding kinoform are non-overlapping. Having discussed the inclusion of two spatially varying lattice parameters (i.e. lattice orientation and lattice spacing) next we suggest the inclusion of a third parameter i.e. fill-factor. A uniform increase or decrease in fill-factor can easily be achieved during the experimental process of recording of structure by controlling the total exposure dose. In addition, we have recently shown that a space dependent variation in fill-factor is possible by doing a coherent sum of two wave-fields of similar kind[11,13]. But, such a coherent sum is not applicable when both the fields are exactly same. Instead, one need to modulate the phase mask to modify the local diffraction efficiency. This modulation in local diffraction efficiency leads to control over local intensity of interference profile providing the desired variation of fill-factor. This method is described in detail by Davis et al.[22]. Thus, we see that it is possible to encode a lattice wave-field with varying fill-factor as well in addition to the lattice-orientation and lattice-spacing parameters by using a phase only mask.

For the experimental demonstration, we use a diode laser (Toptica-BlueMode, Germany) emitting 50mW at 405nm wavelength, a 20x microscope objective and lenses with focal lengths of 135 mm (for collimation), f1 = 500 mm and f2 = 135 mm (for 4f Fourier filter setup), and a phase-only SLM (Holoeye-LETO, Germany), which is a reflective liquid crystal on silicon microdisplay with 1920x1080 pixels and 6.4μm pixel pitch. Fig. 4(a) shows the schematic of experimental setup. The numerically computed kinoforms or phase masks were electronically displayed on the SLM. The incident collimated laser beam gets modulated as per the displayed phase mask and generates multiple diffracted beams. There is a non-zero reflection at SLM-air interface which gives rise to the unwanted zero order diffraction term in the Fourier transform plane. A Fourier filter, designed by making use of numerically obtained Fourier transform profile, was used in the Fourier plane to block all but the desired first order diffracted beams. The second lens of 4f setup helps all the beams to come together and interfere to form the desired lattice structure in the imaging plane. The ratio of focal lengths i.e. f1/f2 controls the demagnification factor in this arrangement. So, f1 = f2 = 500mm was used with a CMOS camera (UI-1488LE-M IDS, Germany) to record the generated interference pattern and f1 = 500mm, f2 = 135mm combination was used for the microfabrication of this spatially varying lattice. For this microfabrication, we used a positive photoresist (AZ1518) which was spin-coated (using Spin-NXG-P1, Apex Instruments, India) on a glass slide at 4000 RPM for 60 seconds to obtain approximately 2μm thick layer. The photoresist film was then prebaked on a hot-plate (EchoTherm-HP50, Torrey Pines Scientific, USA) for 50 seconds at 100°C. Pre-baked sample was exposed to the interference pattern for 9 seconds. The exposed sample was then developed in the developer solution (AZ726 MIF) for 60 seconds. Developed sample was imaged (Fig. 4(b)) by a benchtop scanning electron microscope (SEM) (JEOL NeoScope-JCM6000, Japan).

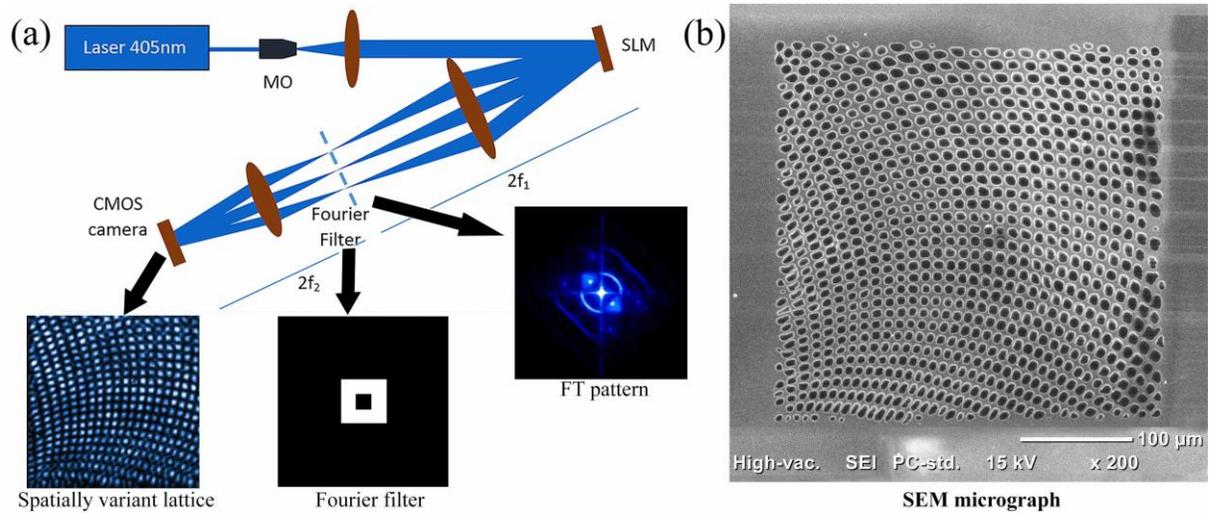

FIG. 4: (Color online) Experimental setup and obtained the results. MO: microscope objective and SLM: Spatial light modulator (phase only). (a) Schematic of the 4f Fourier filtering based experimental setup. Experimentally obtained Fourier transform irradiance profile, corresponding Fourier filter (where black means zero transmittance and white means fully transparent) and resultant irradiance profile captured by CMOS camera are also shown. (b) SEM micrograph of the structure recorded in the photoresist medium.

The method being a general one could easily be extended to more complex lattice structures as well. Although not demonstrated here, instead of having a uniform amplitude for all the terms in Eq. 8, it is possible to spatially vary the individual amplitude terms to vary the unit cell geometry throughout the lattice[13]. Only limitation is in terms of the requirement of non-overlapping diffraction terms of the kinoform which restricts this method to a spatially varying lattice where for a given direction, maximum frequency does not exceed the minimum frequency by a factor of 2. This is not a serious limitation as this could actually be overcome by firstly, dividing the structure in multiple segments where each segment satisfies the frequency variation requirements and secondly, fabricating the structure by making use of stitching with a precision translation stage and multiple exposures. Thus, the method could be very useful in fabrication of a wide variety of photonic crystal elements e.g. Lüneburg lens[23], beam splitter[24], high Q defects[25] etc.

M.K. wishes to acknowledge the Council of Scientific & Industrial Research (CSIR), India for providing the senior research fellowship during the present work. J.J. wishes to acknowledge the research grants supports from DeitY and DRDO, Govt. of India.